\documentclass[journal]{IEEEtran}
\usepackage[none]{hyphenat}
\usepackage{subfig}
\usepackage{xcolor}
\usepackage[draft]{todonotes}   
\usepackage{multirow}
\usepackage{eurosym}
\usepackage{graphicx}
\usepackage{epstopdf}
\usepackage{caption}
\usepackage{booktabs}
\usepackage{amsmath}
\newcommand{\head}[1]{\textnormal{\textbf{#1}}}

\begin{document}
%
\title{From competitive to cooperative aggregation, shifting from auctions to fairs}
\title{Double-side aggregation for e-commerce: \\the fair model}
\title{Aggregating buyers and sellers for e-commerce: \\ how demand and supply meet in fairs}
\title{Fairs for E-commerce: \\the Benefits of Aggregating Buyers and Sellers}
%
%
%

\author{Pierluigi Gallo,~
        Francesco Randazzo,~ 
        and~Ignazio Gallo
    \thanks{P. Gallo is with DEIM, Universit\`{a} di Palermo, viale delle Scienze ed. 9, 90128 Palermo, Italy, e-mail: pierluigi.gallo@unipa.it}
      \thanks{I. Gallo is with Universit\`{a} dell'Insubria, Via Ravasi 2, 21100 Varese, Italy, e-mail: ignazio.gallo@uninsubria.it}
    }

\maketitle
\thispagestyle{empty}

\begin{abstract}
\boldmath
In recent years, many new and interesting models of successful online business have been developed. 
Many of these are based on the competition between users, such as online auctions, where the product price is not fixed and tends to rise. 
Other models, including group-buying, are based on cooperation between users, characterized by a dynamic price of the product that tends to go down.
There is not yet a business model in which both sellers and buyers are grouped in order to negotiate on a specific product or service.
The present study investigates a new extension of the group-buying model, called \textit{fair}, which allows aggregation of  demand  and supply for price optimization, in a cooperative manner.
Additionally, our system also aggregates products and destinations for shipping optimization.
We introduced the following new relevant input parameters in order to implement a double-side aggregation:
(a) \textit{price-quantity curves} provided by the seller;
(b) \textit{waiting time}, that is, the longer buyers wait, the greater discount they get;
(c) \textit{payment time}, which determines if the buyer pays before, during or after receiving the product;
(d) the \textit{distance} between the place where products are available and the place of shipment, provided in advance by the buyer or dynamically suggested by the system. 
To analyze the proposed model we implemented a system prototype and a simulator that allow to study effects of changing some input parameters.
We analyzed the dynamic price model in fairs having one single seller and a combination of selected sellers.
The results are very encouraging and motivate further investigation on this topic.
\end{abstract}
\begin{IEEEkeywords}
auction, aggregation, fair, group buying, social buying
\end{IEEEkeywords}

\section{Introduction}
\label{s:introduction}
\IEEEPARstart{I}{n} recent years, the expansion of e-commerce services has led to the creation of new Internet based business models, including auctions and group buying.
 
Online auctions is becoming very popular both in business-to-business (B2B) and in consumer markets.  
Unlike fixed prices mechanisms (FPMs), which dominated pricing strategies in last decades, online auctions introduce dynamic pricing mechanisms (DPMs) where buyers also dynamically influence the sale price. 
English and Vickrey auctions are the most popular on the Internet, but several different auction schemes exist, some have just appeared recently. 
The most popular online auction site, eBay, adopts English auctions, but allows suppliers to set a limit to the highest price can be offered for the object. 
The price decision mechanism can even be fully assigned to buyers, as it happens in Priceline.com. 
First, buyers propose their own price for flight tickets and hotel rooms; then, sellers decide whether to accept such prices, based on demand and asset availability.

Recently, also Group Buying (GB) business models appeared on the Internet: buyers cluster in groups to obtain discounts on purchasing products and services.
In some cases, this works under the condition that a minimum number of requested items, otherwise potential buyers cannot finalize the purchase. 
GB permits single buyers to get discounts that are normally available only to wholesalers.
Products are displayed on the web site during a time frame, typically called \emph{auction cycle}.
As more buyers join the group, unit price drops down, according to price / quantity function, predetermined by sellers. This can be explicitly revealed to potential buyers or be withheld.

\begin{figure}[t]
	\centering
	\includegraphics[width=0.9\columnwidth]{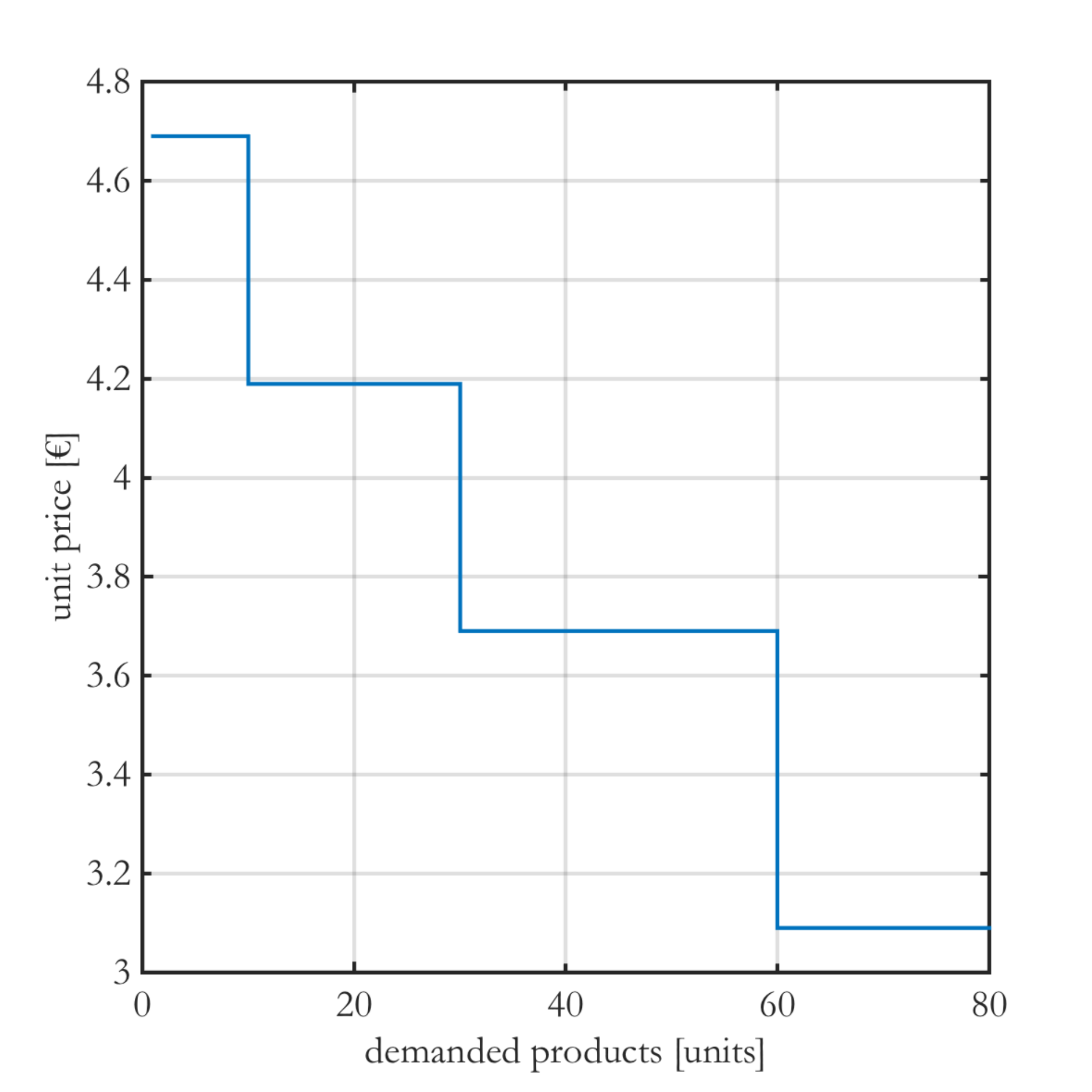}
	\caption{Price trajectory of the unit price of a stack of paper at Staples MondOffice}
	\label{fig_sim}
\end{figure}

Unit price decreasing model is applied not only in group buying, where the demand is aggregated upon several buyers, but also when a single buyer requests more than one object.
A typical price trajectory is reported in \figurename~\ref{fig_sim}, where unit price goes down with demanded quantity, accordingly to 'the more you buy, the more you save' principle. 
Prices move on discrete price levels: from 1 to 9, it is 4.69~\euro, between 10 and 29, it is 4.19~\euro, from 30 to 59 falls to 3.69~\euro and finally it remains 3.09~\euro for higher demands.

A similar pricing trajectory was found on MobShop.com, a pioneer group buying, during an auction cycle \cite{kauffman2001}. 
Mobshop was one of several group-buying sites appeared in the US and European markets in the early 2000s, together with Mercata, CoShopper, and LetsBuyIt.
Despite the brilliant idea of grouping buyers, alone it was not enough to survive the market. 
In facts, all these platforms closed their doors after short time, for bankruptcy or insufficient gains. 
A deep  analysis on the causes of these failures is out of the scope of this paper, however, a common factor was the incapability to capture great discounts and the limited range of product availability\footnote{Further details on these failures are available on online news of the epoch: http://www.wsj.com/articles/SB97951268061999104, http://www.cnet.com/news/group-buying-site-mercata-to-shut-its-doors/}.

Concepts of volume-based discounts and grouping still survive, but recently reappeared in different clothes.
These two aspects reside in contemporary portals: the volume-based discounts is behind the price policy in staples.it, the demand aggregation lays behind  groupon.com.
Groupon uses a fixed price mechanism and does not allow  flexibility in the choice of offered product, that instead is proposed by the platform.
One item is displayed for 24~h as deal-of-the-day, with a high discount rate, even more than 50\%. 
If the minimum number of buyers is reached by 24~h, they receive a redeemable coupon, and the price is debited to their credit cards.

The key aspect, that is surprisingly understudied,  regards the availability of products, the possibility of aggregating \emph{both} buyers and sellers and how this impacts on the unit price.
We deliberately focus on such relevant issue and propose a double-side aggregation that, at the best of our knowledge, is the first methodology which simultaneously take into account buyers and sellers.
Our first contribution is the definition of fairs as key opportunities to aggregate buyers \emph{and} sellers around desired products and services, providing the best trade-off among contrasting aspects. 
A second contribution lays on the definition of a general architecture for fair-based e-commerce. These contributions are organized in the present paper as follows.

First, we provide an overview about related work on e-commerce models, in section \ref{s:related}.
Then, our novel fair-based model and the whole reference architecture are presented, respectively, in sections \ref{s:fair-model} and \ref{s:architecture}. We validated our model implementing a prototype, described in section \ref{s:prototype}. 
Two possible use-cases are reported in section \ref{s:use-cases}, demonstrating benefits of demand and supply aggregation using fairs for products and services. 
We show numerical results about dynamic pricing due to sellers aggregation in section \ref{results}, and how they impact on the whole aggregation of buyers. 
Finally, we draw conclusions in \ref{s:conclusions}, illustrating also directions for future work about fairs.

\section{Related Work}
\label{s:related}
Several studies have explored the online GB phenomenon. The DPM used at the GB sites was compared with the traditional FPM.
In some cases it was found that the DPM is equivalent to optimal FPM \cite{anand2003, chen2007}. 
Further results show that the DPM used in GB has better performance than FPM in four scenarios. 

The first scenario is under uncertain demand regime \cite{anand2003, chen2004}. 
Merchants can set non-linear price-quantity schedules that optimize revenues under each demand regime. 
As better discussed later, our system fits the uncertain demand regime, because it leverages aggregation according to several parameters, including waiting time, payment time, expected savings, etc. 

The second scenario concerns production postponement in combining economies of scale \cite{anand2003,chen2007}. 
This situation emerges when production can be tailored to satisfy the revealed demand information, and economies of scale allow the optimal exploitation of any information that outperforms the FPM.
This also holds in our case, we have flexibility in time and therefore production can be tailored to the product request.

In the third scenario, sellers are risk-seekers who want to expand in a market with new products. 
Risk-seeking merchants desire to sell higher volumes of products, even facing lower unit prices.
This permits to attract more buyers which are price-driven. 

The fourth scenario entails the presence of a greater low-valuation demand than a high-valuation demand \cite{chen2010}. 
Hence, fewer buyers purchase at a higher price, and more buyers exist when the price is lower. In this manner, a seller can gain more benefits by adopting the GB mechanism.

Although these studies provide merchants the tools to get the optimal price curve for GB, they remain theoretical and mainly based on simulations.

Buyers' demand is aggregated during a \emph{waiting time}, which influences GB performances. 
Effects of waiting time on the financial return  are studied in \cite{sharif2009}, according a novel empirical analysis of GB mechanisms. 
This work clearly demonstrates the presence of a trade-off between different performance factors.

Several studies explored group formation mechanism through simulation analysis and modeling.
The earliest GB websites faced difficulties aggregating a sufficient number of buyers with similar purchasing interests \cite{kauffman2001, kauffman2002}.
Afterwards, different strategies were used to improve grouping and in \cite{hyodo2003} it was proposed to arrange buyers in different GB websites.
Authors of \cite{li2010} introduce the concept of Combinatorial Coalition Formation (CCF), which allows buyers to announce reserve prices for combinations of products.
These reserve prices, along with the sellers' price / quantity curves for each product, are used to determine the formation of GB for each product.

In \cite{mastuo2002,mastuo2004} it was proposed the use of a decision support system based on buyer preferences. 
A volume discount mechanism based on the seller's reservation price and the payment adjustment value was the approach used in \cite{mastuo2009}.
GB was proposed also for a whole category of products instead a single one \cite{yamamoto2001} and buyers' web browsing history was used to recommend GB products \cite{chen2012b}.

In addition to consumers’ grouping, also cooperation mechanisms for sellers have appeared in literature. 
A cooperation mechanism among sellers based on exchanging goods in an agent-mediated electronic market system \cite{ito2002a, ito2002b}.
To obtain a higher GB success rate, pricing agents recommend to sellers adequate prices, based on the past buying and selling history data \cite{lee2002}.
Even security has been taken into account and a mechanism for mitigating security risks during GB transactions was proposed in \cite{lee2013}. The authors introduce of a mediator GB server to secure and monitor transactions. 
This helps buyers and sellers to negotiate through a secure channel.

Some studies also investigated the effect that GB mechanism produces on buyers' behavior.
Potential buyers are influenced more by their friends than by marketers \cite{lai2007}.
In our model buyers can have shopping interactions through existing social media, so they can share information about products or services.

 In their analysis about the MobShop website, interesting effects were observed \cite{kauffman2001}.
 Consumers purchase more in:
 large-sized groups (positive reinforcement participation); 
 close to the time when prices drop (price drop effect),
 when the end of an auction cycle is approaching (cycle-ending effect).
 
To take advantage of positive participation 
 three incentive mechanisms where suggested, based on time,  quantity and sequence \cite{lai2004, lai2006}.
 A time-based incentive mechanism can be used to encourage buyers to join the GB auction in its early days by offering an extra participation discount. 
 So the earlier a buyer joins, the higher will be the offered discount.
A quantity-based incentive mechanism helps stimulating buyers to purchase more than planned offering extra discounts on the size of a single order. 
With a sequence-based incentive mechanism, the buyers' times of arrival in the system are compared and discount depends on the sequence of arrivals.
The earlier you join, the less you pay.

However, as already mentioned in section \ref{s:introduction}, most of earlier GB websites have terminated operations. 
An analysis of this early failures is reported in \cite{kauffman2002}, attributing them to:
(1) long GB auction cycles, that hindered buying decision; 
(2) complex GB models, as perceived by buyers;
(3) low transaction volumes, determining small discounts for buyers. It was difficult for earlier GB websites to compete with large retailers from a price perspective.

As discussed in following sections, our model overcomes such negative aspects by introducing:
aggregation of both demand \emph{and} supply for price formation,
aggregation of products and destinations for shipping optimization, 
price prediction,
integration (rather than direct competition) with large e-commerce platforms.

\section{The fair aggregation model}
\label{s:fair-model}
We propose the definition of a new business model in which both sellers and buyers aggregate for negotiating  products or service: \emph{fairs}. 
In their traditional standard form, fairs are organized as gatherings for selling goods, sometimes as exhibitions. 
Fairs include timing concepts (being periodically held), customer aggregation (people meet to discuss about their shopping goals), seller aggregation (sellers mount their booths close each other to permit better product comparison) and amusement, since they provide entertainment.
All these aspects, and even more, are maintained in our fairs, whose architecture is composed by several elements and  parameters, centered on aggregation.

Analogously to standard fairs, which use limited geographic areas to display goods, our fairs provide an aggregated supply coming from multiple sellers in a reduced virtual area (one single platform). 
This increments the attraction mass for potential buyers, because they can look for a wider range of products, services, and sellers as well as their comparison and review (our system integrates with existing social networks).
Furthermore, our system hides the complex optimization procedures behind the optimized unit price quest to buyers, which go through a very simple workflow. 

Fairs work like product-oriented social communities that  aggregate demands at the buyer side and offers at the seller side, with new modalities in the online business world. 
Our system is able to deal also with both goods and services. 
In what follows, for the sake of simplicity, when we refer to goods, buyers and sellers we implicitly include also services, providers and customers.

Buyers' orders are composed according to a large-scale cooperative aggregation that takes into account buyers' location (for better shipping aggregation) and shopping interests (for advanced product selection).
 
The system aggregates orders depending on several buyer-side requirements, including the maximum waiting time and the expected discount.
This basic concept is in common with GB: the more products of the same kind are requested in a single order, the more prices fall down due to scale economy. 
However, unlike standard GB, we can offer better prices even without registered sellers. This can be obtained with our multiple-sided aggregation vision.

High-quantity orders can be achieved with large numbers of participants to the fair. 
Buyers can join the fair requesting a single item or even multiple ones. 
To obtain effective aggregation goals, the system stimulates positive behaviors by providing incentives in terms of applied discounts and promoting social interactions.

Unlike GB, our fair-based e-commerce system also aggregates sellers, looking for the best trade-off among contrasting requirements at the seller and buyer sides: waiting and payment time, price/quantity curves, volumes of available goods and their positions, as well as shipping trajectories.

In our model, seller-side aggregation is both competitive and cooperative. 
The competition among sellers is due to their need to prevail in order to be selected by the system and be called on supplying goods.
Given the (dynamic) demand related to the group of buyers in the fair, and given a list of sellers and their supply conditions, our system dynamically orders sellers by their quality index satisfying buyers' requirements. 

It may happen the aggregated demand of goods cannot be satisfied by a single seller, for the time required. 
Solutions to face such issues are  described in section \ref{s:use-cases}.

Under these circumstances, the demand of the whole fair is partially satisfied by the best seller for that specific fair, then by the second in the list, and so on till the order is completely satisfied. 
Sellers aggregation is cooperative in the sense they collaborate with the system to fulfill the whole supply, but no inter-seller cooperation is neither requested,nor expected, as competing sellers are not supposed to actively cooperate each other.

Fairs are defined given a set of buyers, a set of sellers and their requirements in terms of buying and selling conditions. 
Fairs aggregate demand and supply of a given set of aggregated buyers and sellers. 
Buyers spontaneously aggregate according to their wished product or service, sellers are systematically aggregated by the fair management system.

The system considers sellers that explicitly registered  into the system (so they provided extra information such as price/quantity curves) and includes also those that offer the requested product on integrated e-commerce portals such as Amazon, e-Bay, and Groupon. 
In this case, the system acts as a reseller.

\section{Fair-based sytem architecture}
\label{s:architecture}

\begin{figure*}[tbp]
	\begin{center}
		\includegraphics[width=\textwidth]{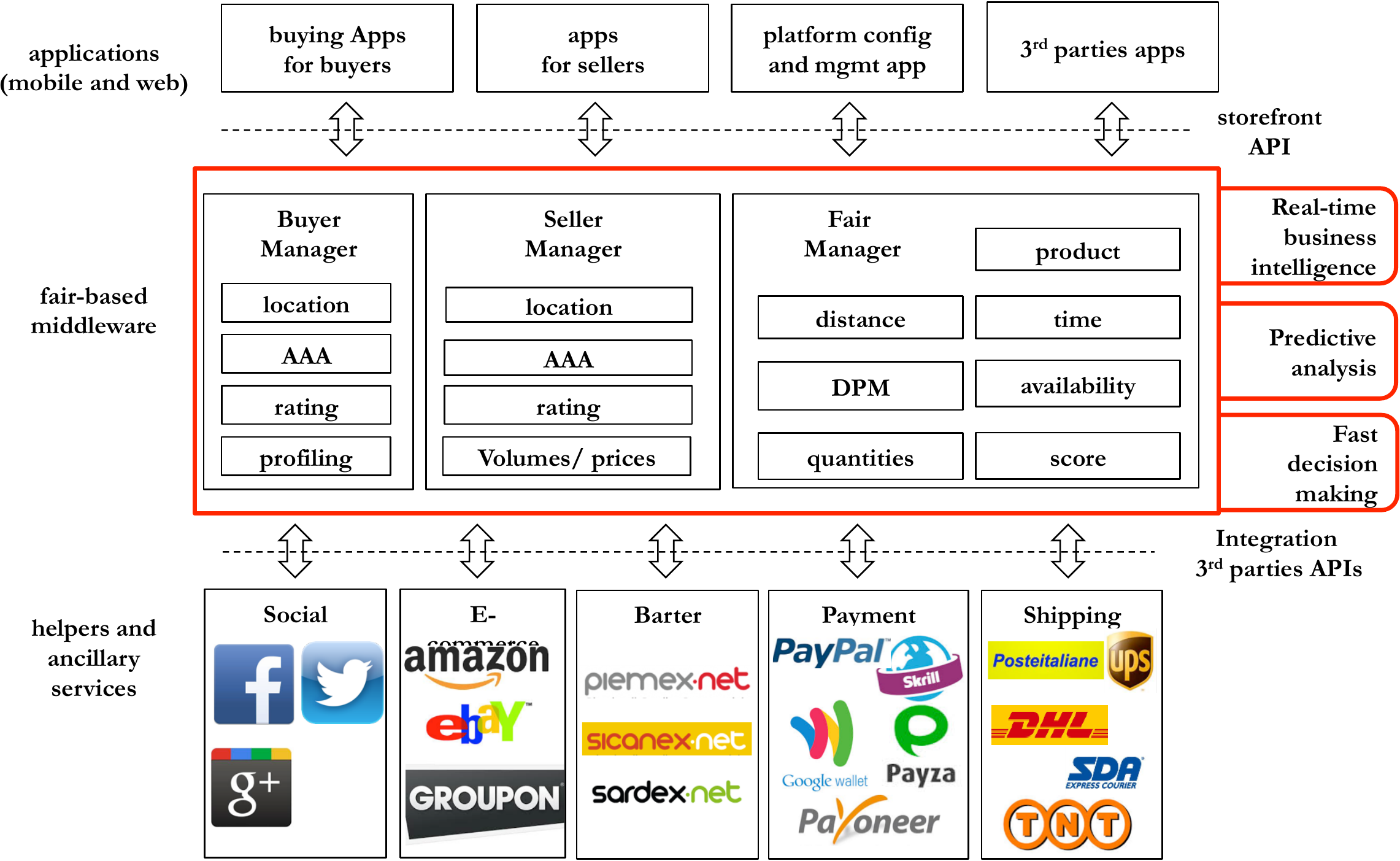}
		\caption{The fair-based system architecture and relevant relationships with external exhisting tools and frameworks.}
	    \label{f:architecture}
	\end{center}
\end{figure*}

Our modular system architecture is reported in fig. \ref{f:architecture}, where several modules handle, configure and manage all actors participating in our fairs.

The \emph{buyer manager} module handles buyers since their authentication into the system, dealing with their demands and profiling. 
Sellers are handled by the \emph{seller manager}, which considers also price/volume curves, location of products or where services can be consumed.
The \emph{fair manager} is a key module that handles most of the fair workflow: groups formation,  DPM handling, seller(s) selection, payment and shipment.
This runs a fair management algorithm that aggregates demands and supply, decides amount(s) paid to seller(s) and how they must be shared between buyers. 
Even if aggregated into the same fair, prices paid by buyers are, in general, heterogeneous, because they are a function also of buyers' fidelity index.

In addition to modules, the middleware exposes open interfaces to develop web-based and mobile applications dedicated to buyers, sellers and platform managers. 
This API permits to integrate our advanced aggregation capabilities into third-parties applications. 
Our system is in the form of a middleware, which relies on existing web services for social interaction.
Buyers interact on social media, rise the interest about fairs and can involve other potential buyers.
 
Payments and shipping services are also externalized to third parties.
However, our middleware provides an  aggregation-based optimization that results transparent to these parties.
Such details are provided to shipment services hiding the complex optimization.
Products and destinations are also aggregated, and the fair management module centrally decides quantities, locations  and shipping sources/destinations in order to minimize shipping costs. 
Users only asynchronously proceed with the final payment and shipping.

The fair algorithm provides also \emph{price predictions} depending on the number of aggregated buyers, thus the user at every moment knows the requested amount for his purchase and future amounts expected under successful aggregations. This encourage buyers to be actively involved in the fair (e.g. by invite their friends interested to the same product categories) and periodically check the portal/application. 
This social aspects of e-commerce may also have emotional value on friendship and socialization \cite{bucci2014}.
 
Sellers no longer set one single price per product, but rather provide a monotone non-increasing curve for price/quantities.
This could eventually have an horizontal asymptote depending on production price.

Fairs end their lifecycle when the maximum time has been reached \emph{or} the optimal minimum price for the fair is obtained. 
This optimal prices is defined in eq. \ref{e:minprice} and in case of multiple equivalent minima, the one corresponding to the lowest aggregated volume is chosen.

\begin{equation}
\label{e:minprice}
z^*_{\pi, \gamma}(q)  = min (z_{\pi, \gamma} (q))
\end{equation}

Given price-quantity curves provided by vendors, given the fair ending event as discussed above, the fair management algorithm determines the quantities to be requested to each seller in order to obtain the minimum unit price and satisfy consumers' demand. 

\subsection{Parameters}
\label{ss:parameters}
As already mentioned, the fair model introduces a novel dimension for the trading concept: buyers and sellers are aggregated according to (competing) parameters.

Significant parameters include  \emph{quantity}  $q_{\pi,\beta,\gamma}$. 
This indicates the amount of product $\pi$ requested by buyer $\beta$ in group (our fair) $\gamma$.
Analogously $q_{\pi,\sigma,\gamma}$ indicates the amount of product $\pi$ requested to seller $\sigma$ in the group $\gamma$. 
Additionally, we consider $Q_{\pi,\gamma}$, as the \emph{availability} of product $\pi$ at the seller $\sigma$, therefore $\sum_{\gamma} q_{\pi,\sigma,\gamma} \leq Q_{\pi,\gamma}$ must hold for each seller, considering all running fairs. 

As for timing, $t_{\beta}$, is the \emph{waiting time} before receiving the product and depends on the fair duration and  $t_{\sigma}$ is the \emph{payment time}, that takes care if the buyer pays before, during or after receiving the product.

From the positioning point of view, our system keeps position of products and the history of buyers' positions, respectively  $\boldsymbol{p}_\beta = (x_\beta,y_\beta)$ and $\boldsymbol{p}_{\sigma} = (x_\sigma,y_\sigma)$.
Buyers that ask for shipping in heterogeneous places, may obtain benefits if they regularly attend common places. The system correlates buyers' positions and provides  suggestions for shipment aggregation. 
As for example, different buyers that periodically attend the same place (e.g. school, office, etc.), can receive their parcel when they go there. 
This permits to obtain reducing shipping costs. 
Position of buyers is taken by the mean of a simple positioning tool integrate into the web service and mobile application.

Incentives are given to buyers and sellers depending on their fidelity scores $\phi_{\beta}$ and $\phi_{\sigma}$. 
As for buyers, these depend on their previous interaction with the system: the number of purchases, payment time (the earlier the better), and number of fair-related actions in social networks. 
Sellers are scored accordingly to the shape of their price/quantity curve (in case of registered sellers) and about the reliability of data they provide (shipping time, availability, etc.).

As for prices, the \emph{unit price} to be paid by the buyer $\beta$ for the product $\pi$ is $z_{\pi,\beta}$.
Analogously the price requested by the seller $\sigma$, for the same product, is  $z_{\pi,\sigma}$. 
Demand and supply meet in fairs with a unit price for product $pi$ in the fair $\gamma$, which is indicated as $z_{\pi,\gamma}$ and can eventually differ from both previously mentioned prices. 

Additionally, the fair manager business model, includes a small revenue upon fairs that reached the end of their lifecycle.
The gain for the fair manager is the difference between the two sums in eq. \ref{e:revenue}.
\begin{equation}
\label{e:revenue}
\sum_{\beta \in \mathcal{B}_\gamma} z_{\pi, \beta}  \geq \sum_{\sigma \in \mathcal{S}_\gamma} z_{\pi, \sigma}
\end{equation}

\begin{figure*}[tbp]
	\begin{center}
		\includegraphics[width=\textwidth]{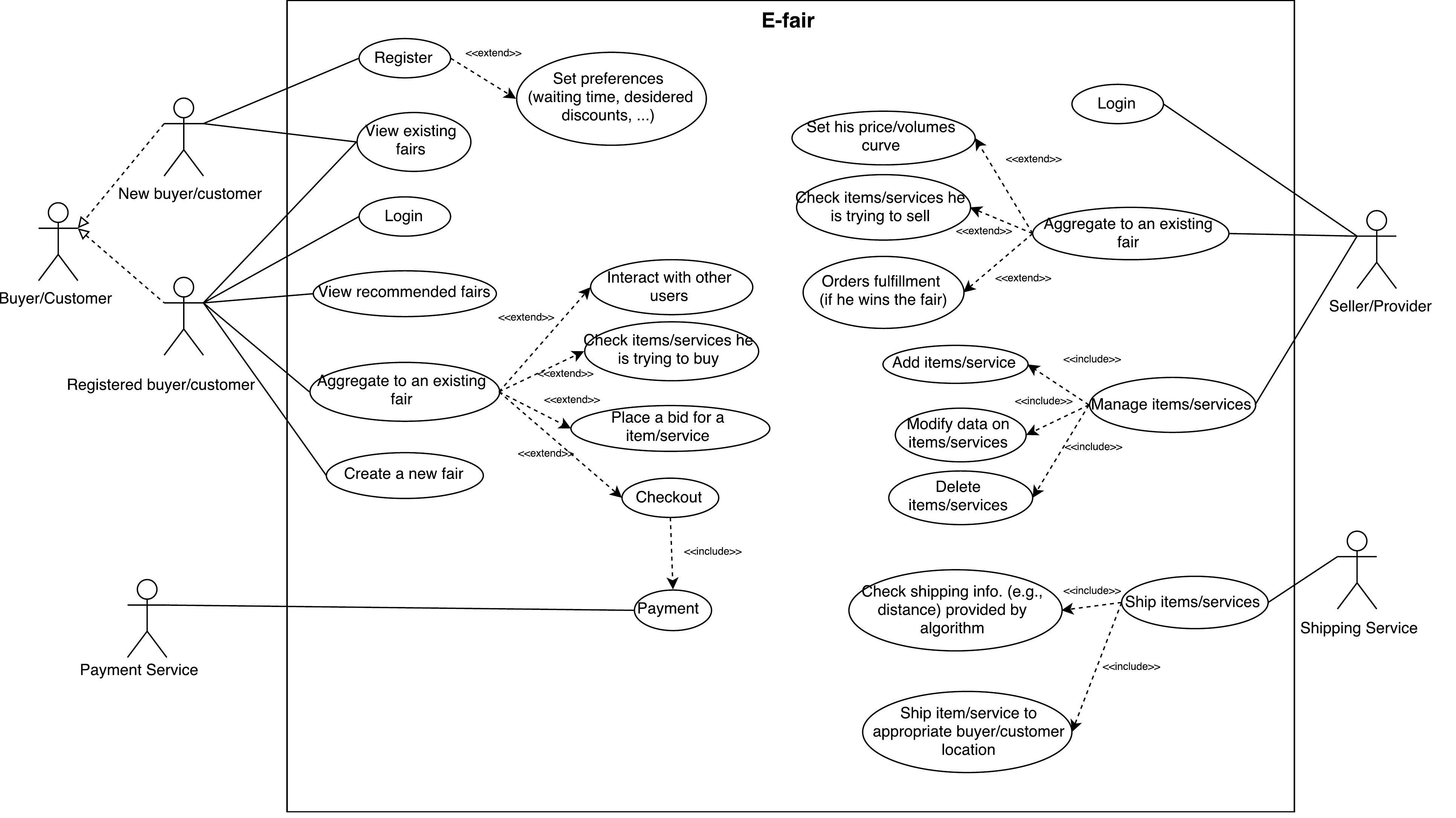}
		\caption{The UML use case diagram that shows the main functions of a fair-based system.}
		\label{f:use_case}
	\end{center}
\end{figure*}


\section{System Prototype}
\label{s:prototype}
We designed and implemented a first system prototype in order to validate the feasibility of our proposed fair model over a real e-commerce framework.

From the available open-source platforms, we analyzed Zen Cart, OpenCart, and Magento.
We finally selected OpenCart \cite{opencart}, as a basis to implement our idea, because of its integration with a large number of services offered by third-parties. 
OpenCart is free open source ecommerce platform for online merchants and it is available under the GNU GPL. 
The software is written in PHP with a MySQL database by default. 
OpenCart offers a number of features permitting to deal with unlimited products, manufacturers and multiple shops. 
Despite it comes with unlimited categories, some extensions have been necessary to support services rather than just products. 
The framework natively integrates several payment systems and shipping methods, eventually configurable for different geographic area, which is a key functionality of the proposed architecture.
The extensibility of the system is due to its modularity and in a dedicated e-store  more than 9000 modules and themes are already available. 

In our first prototype, we extended OpenCart functionalities adding the following capabilities:
(i) handling DPM by the means of discount factors, as discussed in sect. \ref{ss:parameters};
(ii) pricing updates depend on both the requested quantity and the product availability in stock.
Prices, quantities, and other relevant parameters have been implemented as multiplicative factors;
(iv) user's position is taken and stored in the database, for user profiling and shipping optimization, by the mean of an Ajaxcall;
(v) sellers provide their multi-dimensional pricing strategies.

\section{Use cases}
\label{s:use-cases}
\figurename~\ref{f:use_case} shows the UML use case diagram for a fair-based system.
Main system actors involved in a fair are buyers (or customers) and sellers (or providers), as well as payment and shipping services.

We describe the whole workflow for fair management by the means of an example. 
Let's consider having a buyer that wants to buy a smartphone, which can be found online at a price of 100~CUs.
After logging into the system using private credentials, buyers check if existing fairs are already running about the desired object (category and model). 
If there is no  running fair for the demanded product, the buyer launches a new one.
Opening the fair, he describes the product category and  
 provides parameters about desired discount rate, time constraints, payment constraints.
In case the desired product is not available among those supplied by registered sellers, the platform interrogates external e-commerce portals (e-bay, groupon, amazon, etc.), in order to return anyway an answer to the questing buyer.
This avoids frustrating potential users with lack of products, which can be probable during the starting period, when the number of registered sellers and products is not high.
Of course, having no price/quantity curves available, the only benefits would be those of shipping aggregation. 

As for sellers, they access a dedicated section where they define the price/quantity curve by providing its parameters or through a table.
When the fair goes to its end, all buyers involved receive a notification and the final price is communicated. 
Payment is requested to those buyers that did not do it before.
The system optimizes shipping services after determining quantities to be requested to each seller and computed the \emph{best shipping routing} between sellers and buyers.

Fairs deal also with services and in this case we refer to customers and providers.
\emph{Entertainment services} (tickets for museums, cruise, flights, ...), \emph{health services} (medical treatments), and \emph{learning services} (private lessons) can be dealt with fairs.
As for providers, extra aggregation models are available, depending on the time the service is consumed.
Both \emph{sequential} and \emph{concurrent} aggregations are possible as customers can use the service concurrently (e.g. in a cinema) or in a sequential manner (e.g. scheduled wellness treatments in a beauty farm).


\section{Results}
\label{results}
In this section we show results obtained by our aggregation mechanism. 
Being aggregation at the buyer side extensively covered in literature, we synthesize the aggregation at the buyer-side with the aggregate number of demanded items, then we focus on aggregation at the seller side.

To validate our aggregation methodology we used both the system prototype described in sect. \ref{s:prototype} and an ad-hoc simulator written in Matlab, to test aggregation over a large number of sellers and buyers.

\subsection{Dynamic pricing model for the single seller}
\begin{figure}[tbp]
	\begin{center}
		\includegraphics[width=\columnwidth]{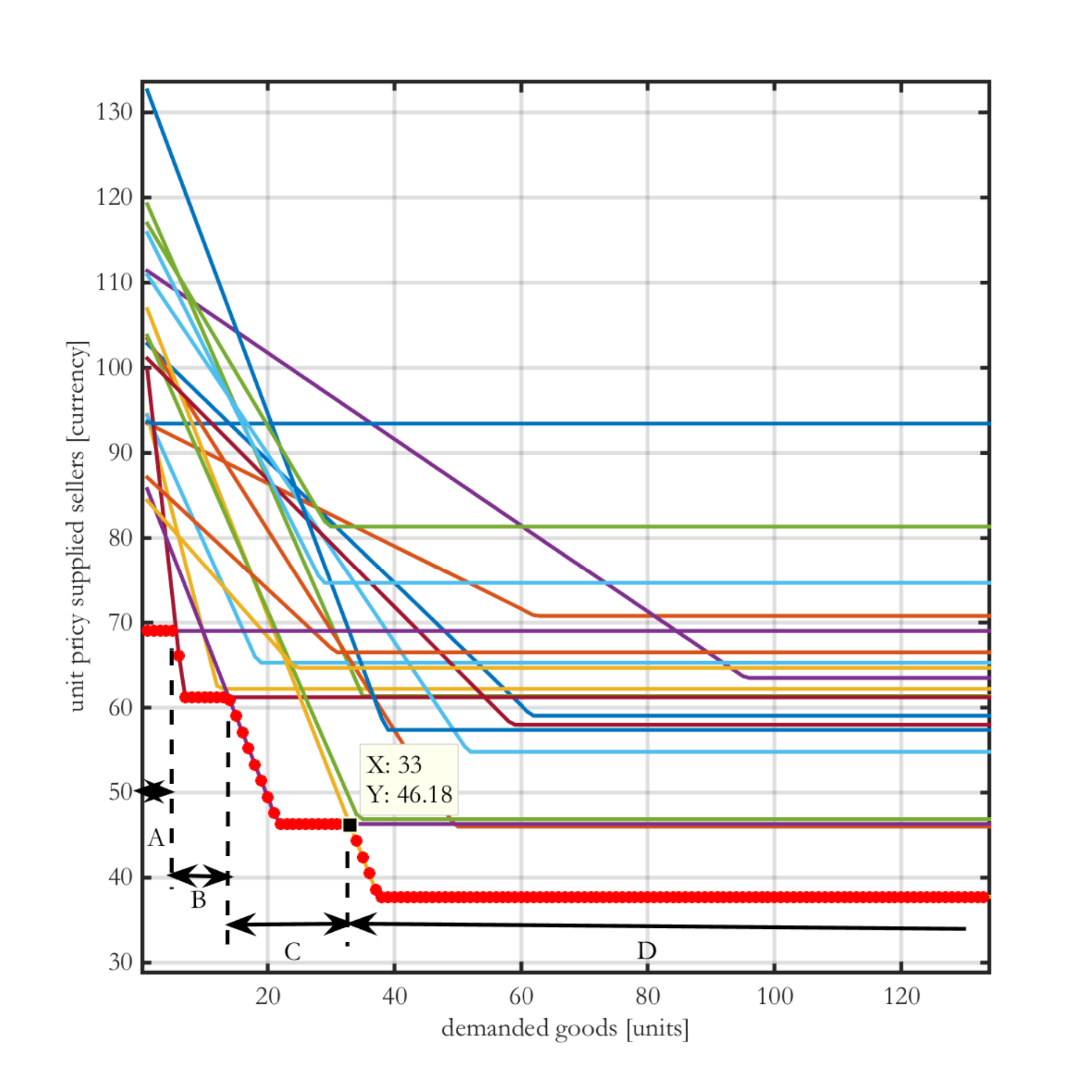}
		\caption{Price/quantity diagrams for a single product. Each line represent the curve for a different seller. The curve defines how the unit price changes while varying the demand.}
		\label{f:prices-curves}
	\end{center}
\end{figure}

In fig. \ref{f:prices-curves}, the dynamic pricing models for single seller are shown.
Among the test campaign we run, we show results using 20 sellers providing as many curves of unit price/quantity, for the same good.

Our aggregation middleware supports whatever shape of dynamic pricing model for sellers, which can provide the price as (continuous) functions of the demanded quantity or as (discrete) two-columns table: quantity and price.
For the sake of simplicity, without lack of generality, we modeled price/quantity curves  as broken lines: they are slopes till a certain number of demanded products, then they turn into plateau.
Despite its simplicity, this model is realistic enough, because slopes model a decrementing price strategies with constant discount rates represented by the angular coefficients.
From a value on, slopes become plateau, modeling the saturation effect due to the fact that selling unit price cannot be lower than production-related costs. 

The lowest broken line in bold red indicates the lower envelope of DPM curves and is the unit price obtained by the fair (all sellers), under the assumption that each seller has infinite supply availability. 
Under this condition, given the number of demanded products, all of them are available (and bought to) a single seller. 
However, depending on the quantity, the most convenient seller can vary. 
As for example, if the requested quantity is in the range 1 to 5 units, the best seller to select is A, from 5 to 13 is B, from 13 to 33 is C and from 33 on is D, as delimited by vertical dashed segments and indicated by black arrows in fig. \ref{f:prices-curves}.

It worths to note that these DPM models are independent each other. 
These are completely defined by three parameters: 
(i) the single-product unit price, when only one product is demanded (close to the intercept value with y-axis),
(ii) the discount rate (the angular coefficient of the slope), 
(iii) the saturation price (the height of the plateau). 
Values used in tests are obtained using distributions reported in table \ref{t:distributions}.
For the sake of generalization, we use generic currency units (CU) instead of \$, \pounds, \euro, etc.

\begin{table}
	\caption{Parameter values used in simulation}
\begin{tabular}{ccc}
	\toprule[1.5pt]
	\head{DPM parameter} & \head{Distribution} & \head{Dist. params}\\
	\midrule
	\multirow{2}{*}{	$single-product~unit~price$} & \multirow{2}{*}{Normal} & $\mu=100~CU$  \\ 
																											  & &$\sigma=20~CU$\\
	\midrule
	\multirow{2}{*}{$\mid discount~rate \mid$} & \multirow{2}{*}{Lognormal} &  $\mu=-2~CU/u$\\
																											 & & $\sigma=2~CU$ \\
    \midrule
	\multirow{2}{*}{$saturation~price$} & \multirow{2}{*}{Normal} & $\mu=60~CU$ \\
																							& & $\sigma=12~CU$\\
	\bottomrule[1.5pt]
\end{tabular}
\caption*{Realistic distribution kind and parameters for DPM for a single seller. These may depend on the category of product and on the elasticity of demand and supply.}
\label{t:distributions}
\end{table}

\subsection{Dynamic pricing model for the whole fair}

\begin{figure*}[tbp]
	\begin{center}
		\begin{minipage}[b][]{\columnwidth}
			\subfloat[]{\includegraphics[width=\columnwidth]{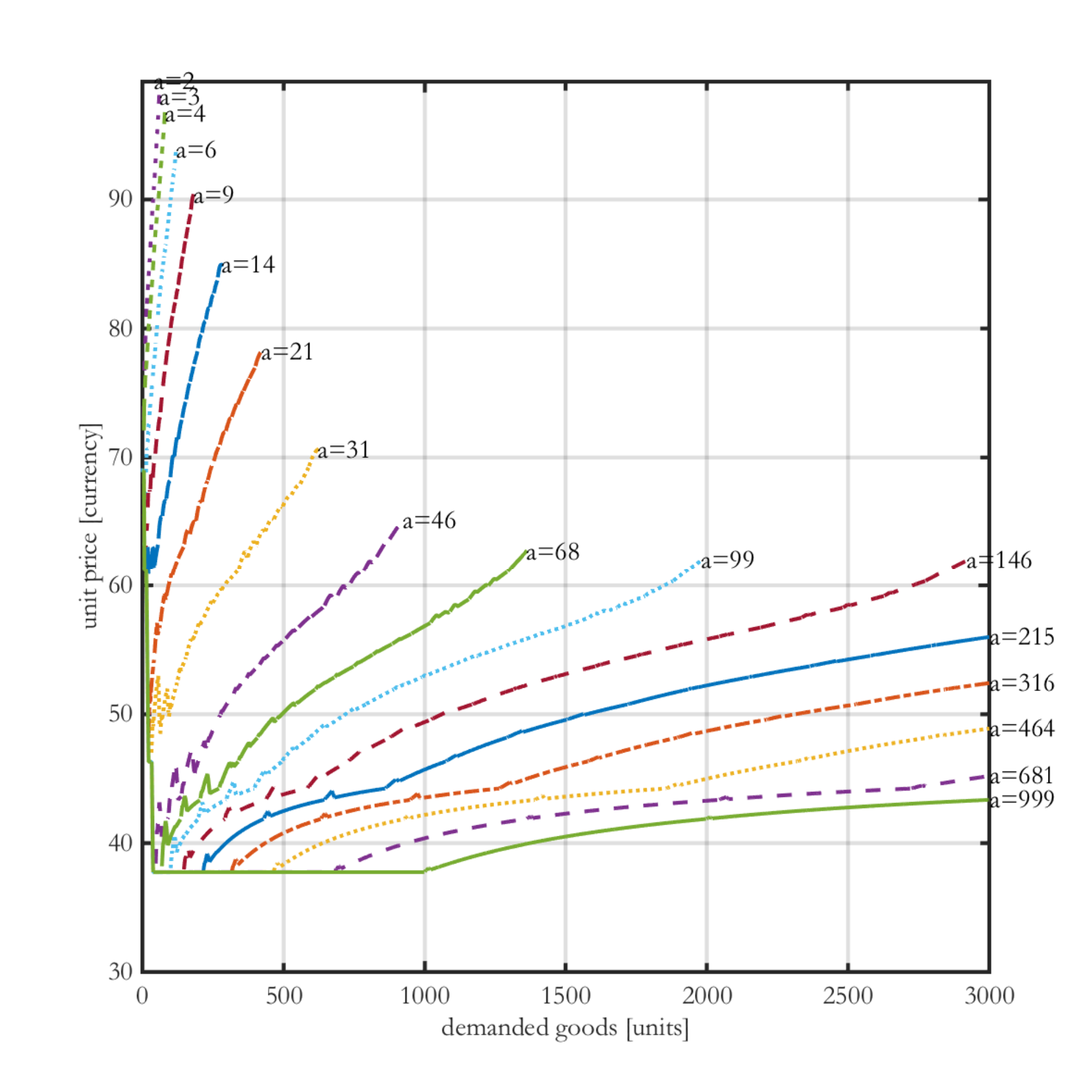}}\quad
			\end{minipage}
			\begin{minipage}[b][5.5cm]{\columnwidth}
			\subfloat[]{\includegraphics[width=\columnwidth]{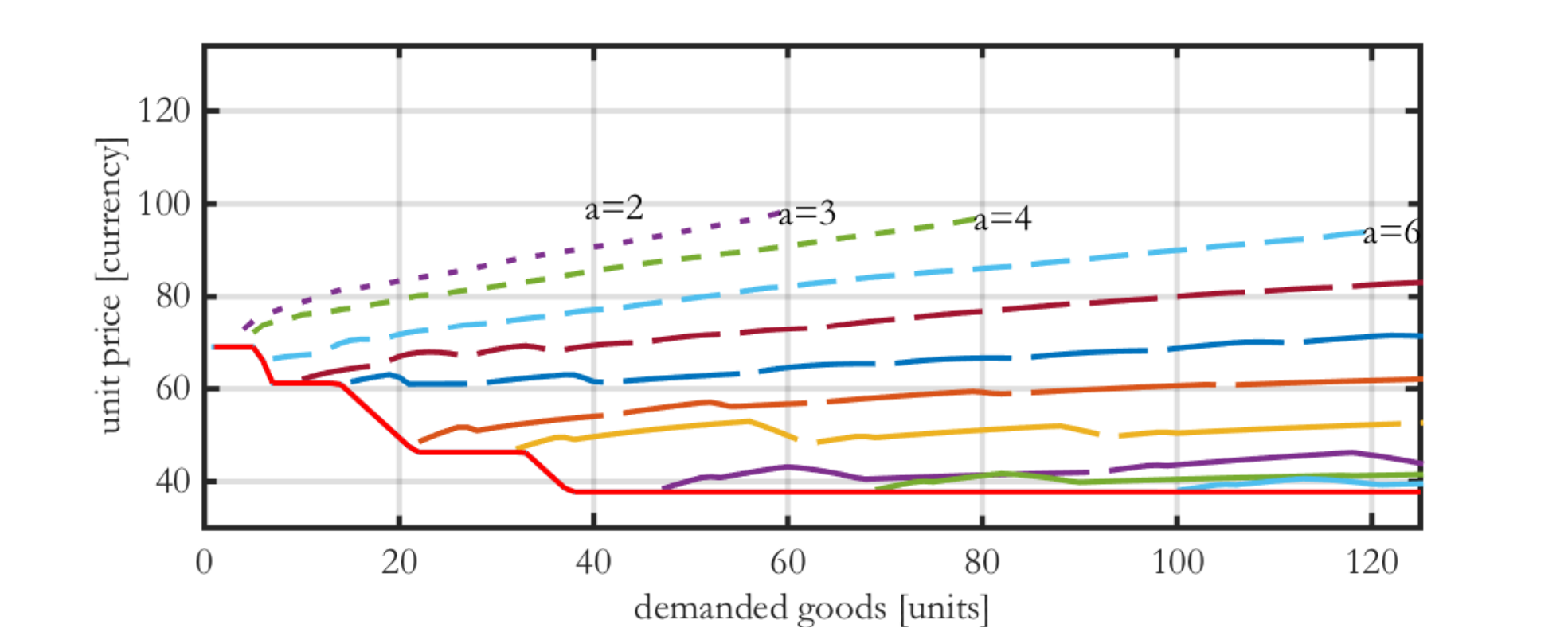}}
			\vspace{0.5cm}
			\subfloat[]{\includegraphics[width=\columnwidth]{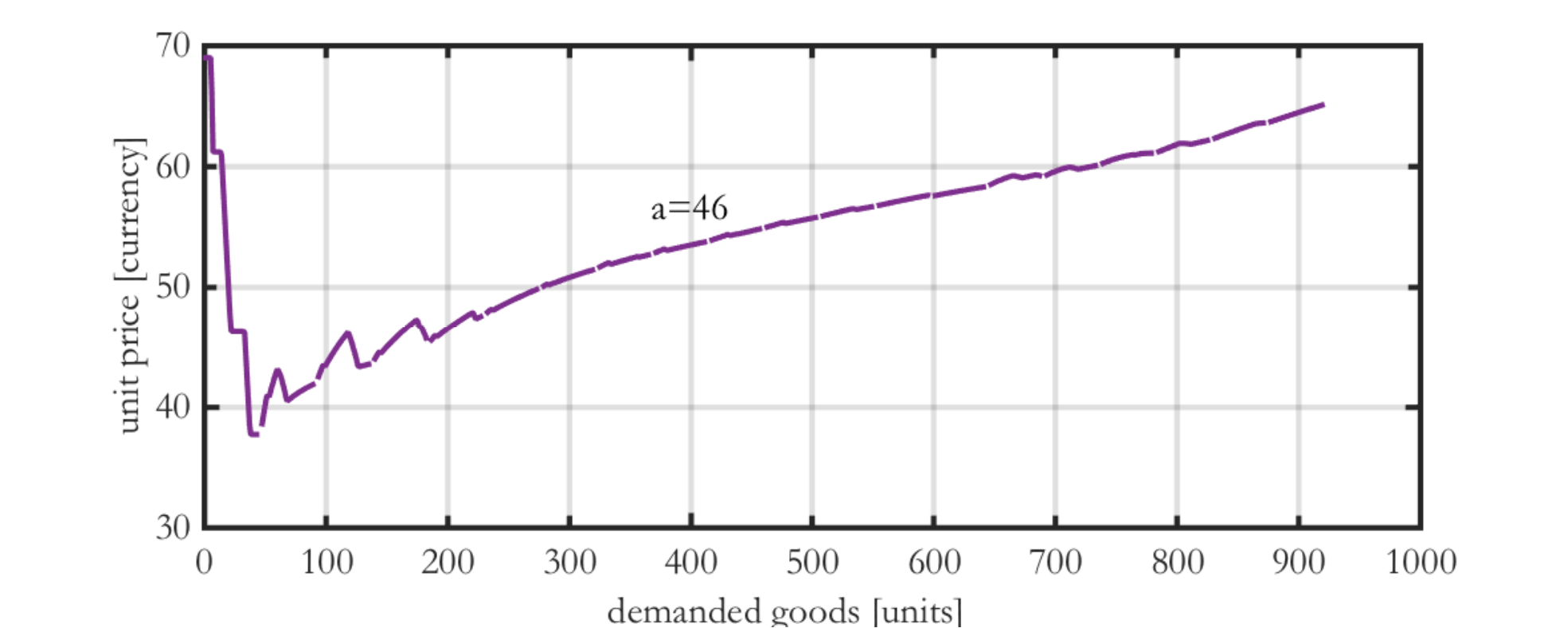}}
			\end{minipage}
			\caption{Dynamic pricing model for the fair, due to sellers aggregation. Each curve shows the trend of the unit price by varying the demanded amount. Curves are (logaritmically) parametrized depending on the number of products available at each seller (a).  The same diagram shown on the left, zoomed to better show its left-most side (b). The price curve when all sellers have 46 products (c).}
			\label{f:prices-dynamics}
	\end{center}
\end{figure*}

Unlike the dynamic pricing model for the single seller, the DPM for the fair can be non-monotone. 
This happens because after the most convenient seller terminates products in stock, the system has to consider the second choice, then the second etc.
The unit price, in such a case is not anymore computed directly on price/quantity curves but using a function of a weighted sum:
\begin{equation}
 z_{\pi,\gamma}(q_\gamma)=l\left( \frac{\sum\limits_{\sigma \in \mathcal{S}_\gamma} q_{\pi, \sigma,\gamma} \cdot z_{\pi,\sigma}}{\sum\limits_{\sigma \in \mathcal{S}_\gamma} q_{\pi, \sigma, \gamma} } \right)
\end{equation}

In our case we used the function $l(x)=x$, therefore directly the weighted sum. 

In fig. \ref{f:prices-dynamics}(a) we present pricing curves obtaining aggregating sellers with a \emph{finite supply availability}, homogeneous over all sellers.
The more products are available at each seller and the more the curve moves towards the lower right corner of figure. 
Different phenomena can be explained by the mean of (b) and (c), both obtained by (a) using different zooms.
In fig. \ref{f:prices-dynamics}(b), it is possible to recognize the same red steps shown in fig. \ref{f:prices-dynamics}. 
These steps are the asymptotic diagram to which fair-based price/quantity curves tend when the number of products available at each seller tends to infinity.
On the other hand, in (c) it is possible to see a the price curve when all sellers have 46 products. It is more evident that this curve has a plateau at its minimum, which represent the optimal aggregation range.
The best fair performances are obtained when the demand is in the range between 38 and 45 products, which is exactly the maximum number of products available at the best seller.

\section{Conclusion and future work}
\label{s:conclusions}
This paper presents a novel aggregation model for both buyers and sellers, paving the road to new possibilities in e-commerce scenarios. 

Some features of our model come from group buying, others from auctions systems but the resulting system lays at equally distance from both, with the new introduction of double-side aggregation.
In systems like e-bay, prices increase due to competing buyers in the auction, while in our system prices become lower due to users' cooperative aggregation. 
However, in a more generic model, which includes also shipping and auxiliary costs,  prices/quantity curves are not monotonically descent with the number of goods. 
This requires the solution of an optimization problem, which is computed by the fair manager module.
The output of this algorithm is expected to provide the minimum unit price and the optimal quantity to be requested to each seller. 

Effects of elastic demand / supply models over the system should be investigated, as well as the case of heterogeneous availabilities at different sellers and how contrasting factors interact each other (e.g. the best price seller for a specific quantity may have more expensive delivery fares).

Despite this work is still preliminary, the encouraging results stimulate further investigations on double-side aggregation according to the fair-based model.
As a future work we consider providing sellers with suggestions for optimal selling strategies and accepting eventual divergences they explicitly force in the system.

\section*{Acknowledgment}
The authors are grateful to Giampiero Tarantino and Davide Taibi for their fruitful discussions and insightful suggestions, and to Giuseppe Spedito for his implementation effort on OpenCart. This work has been partially founded by 7Pixel S.r.l..

\ifCLASSOPTIONcaptionsoff
  \newpage
\fi

 \bibliographystyle{IEEEtran}
\bibliography{references}

\begin{thebibliography}{10}
\providecommand{\url}[1]{#1}
\csname url@samestyle\endcsname
\providecommand{\newblock}{\relax}
\providecommand{\bibinfo}[2]{#2}
\providecommand{\BIBentrySTDinterwordspacing}{\spaceskip=0pt\relax}
\providecommand{\BIBentryALTinterwordstretchfactor}{4}
\providecommand{\BIBentryALTinterwordspacing}{\spaceskip=\fontdimen2\font plus
\BIBentryALTinterwordstretchfactor\fontdimen3\font minus
  \fontdimen4\font\relax}
\providecommand{\BIBforeignlanguage}[2]{{%
\expandafter\ifx\csname l@#1\endcsname\relax
\typeout{** WARNING: IEEEtran.bst: No hyphenation pattern has been}%
\typeout{** loaded for the language `#1'. Using the pattern for}%
\typeout{** the default language instead.}%
\else
\language=\csname l@#1\endcsname
\fi
#2}}
\providecommand{\BIBdecl}{\relax}
\BIBdecl

\bibitem{kauffman2001}
R.~J. Kauffman and B.~Wang, ``New buyers' arrival under dynamic pricing market
  microstructure: The case of group-buying discounts on the internet,''
  \emph{Journal of Management Information Systems}, vol.~18, no.~2, pp.
  157--188, 2001.

\bibitem{anand2003}
\BIBentryALTinterwordspacing
K.~S. Anand and R.~Aron, ``Group buying on the web: A comparison of
  price-discovery mechanisms,'' \emph{Management Science}, vol.~49, no.~11, pp.
  1546--1562, 2003. [Online]. Available:
  \url{http://dx.doi.org/10.1287/mnsc.49.11.1546.20582}
\BIBentrySTDinterwordspacing

\bibitem{chen2007}
J.~Chen, X.~Chen, and X.~Song, ``Comparison of the group-buying auction and the
  fixed pricing mechanism,'' \emph{Decision Support Systems}, vol.~43, no.~2,
  pp. 445--459, 2007.

\bibitem{chen2004}
J.~Chen, Y.~Liu, and S.~Xiping, ``Group-buying online auction and optimal
  inventory policy in uncertain market,'' \emph{Journal of Systems Science and
  Systems Engineering}, vol.~13, no.~2, pp. 202--218, 2004.

\bibitem{chen2010}
J.~Chen, R.~J. Kauffman, Y.~Liu, and X.~Song, ``Segmenting uncertain demand in
  group-buying auctions,'' \emph{Electronic Commerce Research and
  Applications}, vol.~9, no.~2, pp. 126--147, 2010.

\bibitem{sharif2009}
H.~Sharif-Paghaleh, ``Analysis of the waiting time effects on the financial
  return and the order fulfillment in web-based group buying mechanisms,'' in
  \emph{Proceedings of the 2009 IEEE/WIC/ACM International Joint Conference on
  Web Intelligence and Intelligent Agent Technology-Volume 01}.\hskip 1em plus
  0.5em minus 0.4em\relax IEEE Computer Society, 2009, pp. 663--666.

\bibitem{kauffman2002}
R.~J. Kauffman and B.~Wang, ``Bid together, buy together: On the efficacy of
  group-buying business models in internet-based selling,'' pp. 99--137, 2002.

\bibitem{hyodo2003}
M.~Hyodo, T.~Matsuo, and T.~Ito, ``An optimal coalition formation among buyer
  agents based on a genetic algorithm,'' in \emph{Developments in Applied
  Artificial Intelligence}.\hskip 1em plus 0.5em minus 0.4em\relax Springer,
  2003, pp. 759--767.

\bibitem{li2010}
C.~Li, K.~Sycara, and A.~Scheller-Wolf, ``Combinatorial coalition formation for
  multi-item group-buying with heterogeneous customers,'' \emph{Decision
  Support Systems}, vol.~49, no.~1, pp. 1--13, 2010.

\bibitem{mastuo2002}
T.~Matsuo and T.~Ito, ``A decision support system for group buying based on
  buyers' preferences in electronic commerce,'' in \emph{the proceedings of the
  Eleventh World Wide Web International Conference (WWW-2002)}, 2002, pp.
  84--89.

\bibitem{mastuo2004}
------, ``A group formation support system based on substitute goods in group
  buying,'' \emph{Systems and Computers in Japan}, vol.~35, no.~10, pp. 23--31,
  2004.

\bibitem{mastuo2009}
T.~Matsuo, ``A reassuring mechanism design for traders in electronic group
  buying,'' \emph{Applied Artificial Intelligence}, vol.~23, no.~1, pp. 1--15,
  2009.

\bibitem{yamamoto2001}
J.~Yamamoto and K.~Sycara, ``A stable and efficient buyer coalition formation
  scheme for e-marketplaces,'' in \emph{Proceedings of the fifth international
  conference on Autonomous agents}.\hskip 1em plus 0.5em minus 0.4em\relax ACM,
  2001, pp. 576--583.

\bibitem{chen2012b}
T.~Chen, ``Towards convenient customer--driven group--buying: an intelligent
  centralised p2p system,'' \emph{International Journal of Technology
  Intelligence and Planning}, vol.~8, no.~1, pp. 16--31, 2012.

\bibitem{ito2002a}
T.~Ito, H.~Hattori, and T.~Shintani, ``A cooperative exchanging mechanism among
  seller agents for group-based sales,'' \emph{Electronic Commerce Research and
  Applications}, vol.~1, no.~2, pp. 138--149, 2002.

\bibitem{ito2002b}
T.~Ito, H.~Ochi, and T.~Shintani, ``A group-buy protocol based on coalition
  formation for agent-mediated e-commerce,'' \emph{IJCIS}, vol.~3, no.~1, pp.
  11--20, 2002.

\bibitem{lee2002}
Y.~K. Lee, S.~W. Kim, M.~J. Ko, and S.~E. Park, ``Pricing agents for a group
  buying system,'' in \emph{EurAsia-ICT 2002: Information and Communication
  Technology}.\hskip 1em plus 0.5em minus 0.4em\relax Springer, 2002, pp.
  693--700.

\bibitem{lee2013}
J.-S. Lee and K.-S. Lin, ``An innovative electronic group-buying system for
  mobile commerce,'' \emph{Electronic Commerce Research and Applications},
  vol.~12, no.~1, pp. 1--13, 2013.

\bibitem{lai2007}
M.~Lai and C.~Su, ``An empirical test of the effectiveness of communication
  source: A case of group buy,'' in \emph{Proceedings of the 13th Asia Pacific
  Management Conference, Melbourne, Australia}, 2007, pp. 1263--1269.

\bibitem{lai2004}
H.~Lai and Y.~Zhuang, ``Comparing the performance of group-buying models with
  different incentive mechanisms,'' in \emph{Proceedings of the Third Workshop
  on e-Business}, 2004, pp. 1--12.

\bibitem{lai2006}
H.~Lai and Y.-T. Zhuang, ``Comparing the performance of group-buying
  models-time based vs. quantity based extra incentives,'' in \emph{Proceedings
  of the Fourth Workshop on Knowledge Economy and Electronic Commerce}, 2006,
  pp. 81--90.

\bibitem{bucci2014}
L.~Bucci, ``From e-commerce to social commerce: exploring global trends,''
  2014.

\bibitem{opencart}
``Opencart - open source shopping cart solution.''

\end{thebibliography}

\begin{IEEEbiography}[{\includegraphics[width=1in,height=1.25in,clip,keepaspectratio]{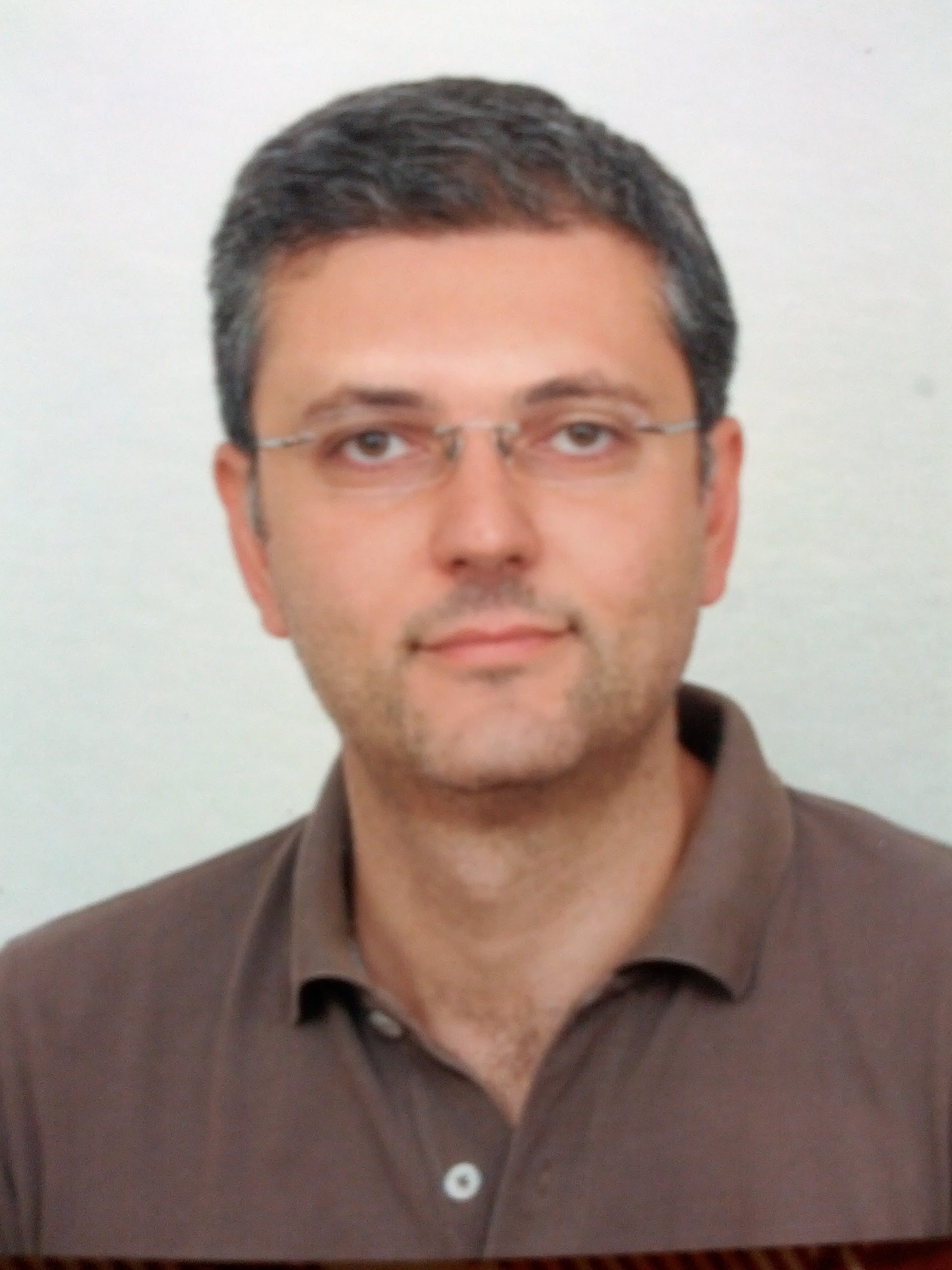}}]{Pierluigi Gallo} has been an Assistant Professor at the University of Palermo since November 2010.
	He graduated with distinction in Electronic Engineering in July 2002 and worked at CRES (Electronic Research Center in Sicily) until 2009. His work there was dedicated to QoS in IP core routers, IPv6 network mobility and Wireless Networks. His research activity has focused on wireless networks at the MAC
	layer and 802.11 extensions, localization based on the time of arrival and cross layer solutions. 
	P. Gallo has contributed to several national and European research projects: ITEA-POLLENS (2001-2003) on a middleware platform for a programmable router; IST ANEMONE (2006-2008) about IPv6 mobility; IST PANLAB II on the infrastructure implementation for federating testbeds; ICT FLAVIA (2010-2013) on Flexible Architecture for Virtualizable future wireless Internet Access; CREW (2013-2014) Cognitive Radio Experimental World; WiSHFUL(2015-) Wireless Software and Hardware platforms for Flexible and Unified radio and network control.
\vspace{-20ex}
\end{IEEEbiography}
\begin{IEEEbiography}[{\includegraphics[width=1in,height=1.25in,clip,keepaspectratio]{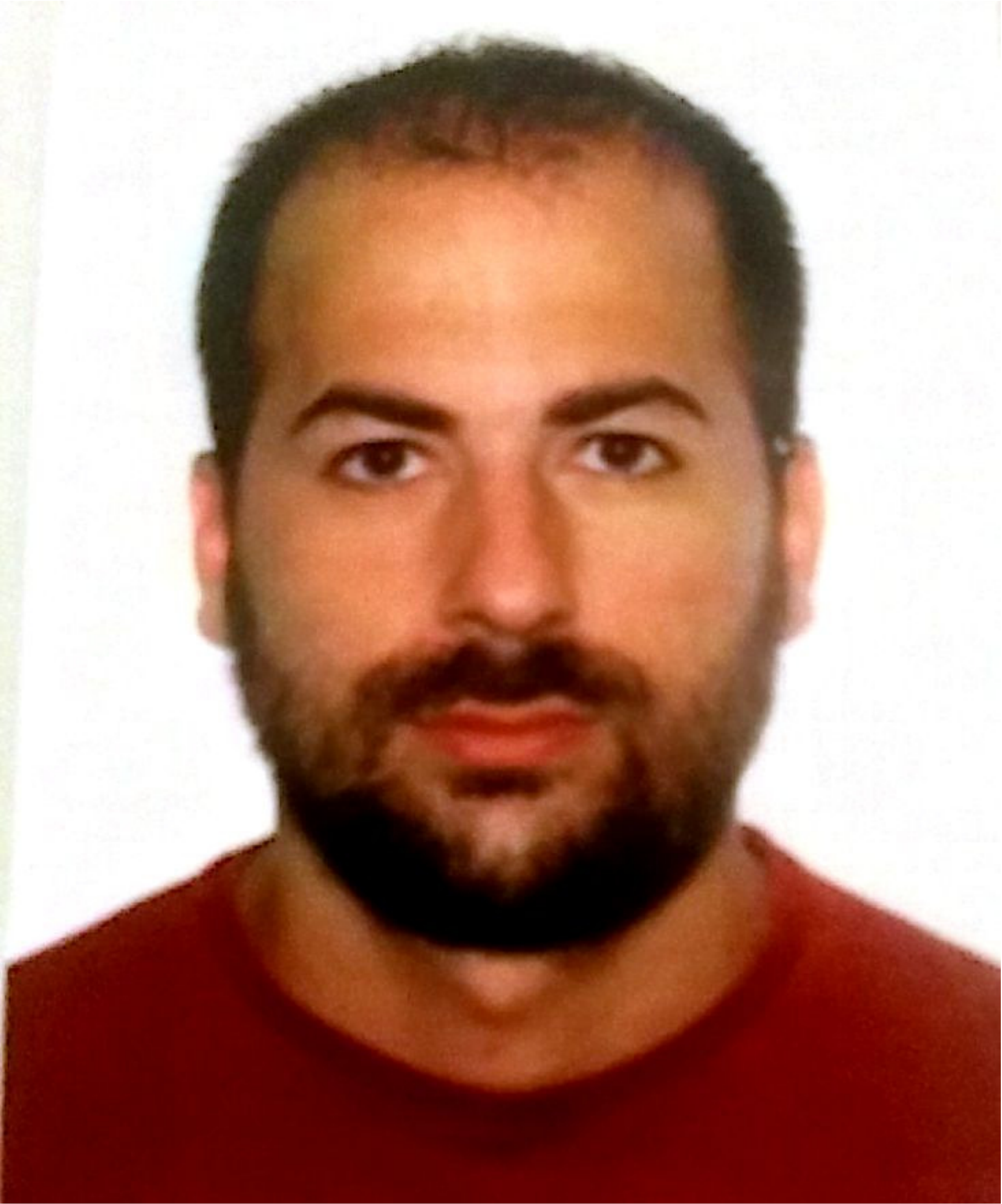}}]{Francesco Randazzo} has been a student at the University of Palermo. He graduated with first-class honours in Telecommunications Engineering in March 2015 and he is working at 7Pixel S.r.l since June 2015. His work is oriented to development of web applications for e-commerce.
\end{IEEEbiography}

\begin{IEEEbiography}[{\includegraphics[width=1in,height=1.25in,clip,keepaspectratio]{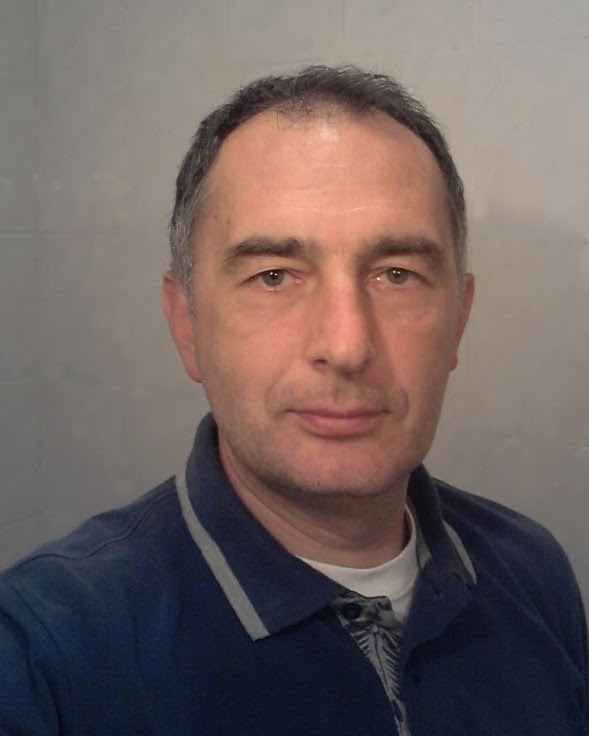}}]{Ignazio Gallo} 
has been an Assistant Professor at the University of Insubria, Varese since 2003.
He received his degree in Computer Science at the University of Milan, Italy, in 1998. 
His research activity has focused on Computer Vision, Image Processing, Pattern Recognition, Neural Computing.
\vspace{-12ex}
\end{IEEEbiography}




\end{document}